\documentclass{llncs}

\usepackage{etex}
\usepackage{delarray}
\usepackage{amssymb}
\usepackage{amsmath}
\usepackage{graphicx}
\usepackage{subfig}
\usepackage[usenames,dvipsnames]{color}
\usepackage{url}

\usepackage{float}

\floatstyle{boxed}
\newfloat{mybox}{htb}{lob}
\floatname{mybox}{Box}
\newsubfloat{mybox}

\usepackage{booktabs}
\usepackage{colortbl}

\usepackage{tikz}
\usetikzlibrary{calc,shapes,snakes,arrows,shadows}

\DeclareCaptionType{copyrightbox}

\usepackage{listings}
\usepackage{caption}
\DeclareCaptionFont{white}{\color{white}}
\DeclareCaptionFormat{listing}{\colorbox[cmyk]{0.43, 0.35, 0.35,0.01}{\parbox{\textwidth}{\centering #1#2#3}}}
\begin{document}

\title{
  Application-tailored Linear Algebra Algorithms:\\ 
  A search-based Approach}

\author{Diego Fabregat-Traver \and 
        Paolo Bientinesi}

\institute{
  AICES, RWTH Aachen, Germany \\
  \email{\{fabregat,pauldj\}@aices.rwth-aachen.de}
}

\maketitle

\begin{abstract}

In this paper, we tackle the problem of 
automatically generating
algorithms for linear algebra operations
by 
taking advantage of problem-specific knowledge. 
In most situations, users possess much more information 
about the problem at hand than what current libraries and 
computing environments accept; 
evidence shows that
if properly exploited, such information leads to 
uncommon/unexpected speedups. 
We introduce a knowledge-aware linear algebra compiler
that allows users to input matrix equations together 
with properties about the operands and the problem itself;
for instance, they can specify that the equation 
is part of a sequence, and how successive instances 
are related to one another. 
The compiler exploits all this information 
to guide the generation of algorithms,
to limit the size of the search space, and
to avoid redundant computations.
We applied the compiler to equations arising 
as part of sensitivity and genome studies; 
the algorithms produced exhibit, respectively, 100- and 1000-fold speedups.

\vspace{2mm}
\end{abstract}

\section{Introduction} 
\label{sec:intro}

The design of efficient application-tailored algorithms 
for matrix operations is an arduous task.
Traditional libraries, typically written in C or Fortran,  
provide a multitude of optimized kernels for 
critical building blocks such as eigenproblems and linear systems, 
but if application-specific knowledge is available,
they lack a mechanism to exploit it.
The burden is thus on the users, who 
have to modify the algorithm and/or the library, 
to tailor the computation to their needs. 
By contrast, 
high-level languages and environments such as Matlab and R~\cite{R}
are designed to deliver solutions automatically, 
without any human intervention.
Unfortunately, this is achieved by giving up optimizations, 
data reuse, and most of knowledge exploitation.
Our goal is to relieve the users from any decision making, 
while still producing solutions that match or even outperform 
those found by human experts.

In this paper, we consider 
equations that involve scalar, 
vector and matrix operands, combined 
through the binary operators ``$+$'' (addition) and ``$*$'' 
(multiplication, used both for scaling and matrix products),
and the unary operators
``$-$'' (negation),
``{\small $T$}'' (transposition), and ``{\small ${-1}$}'' 
(inversion, for scalars and square matrices). 
Equations come with what we refer to as {\em knowledge}: 
Each operand is annotated with a list of zero or 
more properties such as ``square'', ``orthogonal'', ``full rank'',
``symmetric'', ``symmetric positive definite'', ``diagonal'', and so on. 
Additionally, 
we allow operands to be subscripted, 
indicating that the problem has to be solved multiple times.
As an example, Box~\ref{box:input-example} illustrates the description of 
a sequence of linear systems 
that share the same symmetric coefficient matrix: $x_i := A^{-1} b_i$.
\begin{mybox}[!h]
\vspace{3mm}
\begin{verbatim}
               equation          = { equal[ x, times[ inv[A], b ] ] };
   
               operandProperties = {
                  { A,   {``Input'',  ``Matrix'', ``Symmetric''} },
                  { b,   {``Input'',  ``Vector'' } },
                  { x,   {``Output'', ``Vector'' } }
               };
   
               dependencies      = { {A, {}}, {b, {i}}, {x, {i}} };
\end{verbatim}
\caption{Description of the sequence of linear systems $x_i := A^{-1} b_i$,
with a symmetric coefficient matrix. The input, based on the Mathematica language,
includes: the target equation (in prefix notation), the properties of the operands,
and the specific sequence (as dependencies on the corresponding subscript).}
\label{box:input-example}
\end{mybox}

\sloppypar
Simple examples of matrix operations are 
$x := Q^T L \, y$,
$b := (X^T X)^{-1} v$, and
$B_i := A^T_i M^{-1} A_i$; 
in all cases,
the quantities on the right-hand side are known (matrices in capital
letters and vectors in lower case), and the left-hand side has to be
computed. Despite their mathematical simplicity, these equations pose
challenges so significant that even the best tools for linear algebra
produce suboptimal results. 
For instance, Matlab 
uses a cubic---instead of quadratic---algorithm in the first
equation,
incurs possibly critical numerical errors in the second 
one, and
fails to reuse intermediate results---and thus save computation---in the last one. 

Let us take a closer look at $x := Q^T L \, y$: 
Algorithms~\ref{alg:matlab-qly} and~\ref{alg:compiler-qly}
display two alternative ways of computing $x$.
In the left algorithm, the one used by Matlab, 
the input equation is decomposed into a {\sc gemm} 
(matrix-matrix multiplication), followed by a 
{\sc gemv} (matrix-vector multiplication), for a total of $O(n^3)$ 
floating point operations (flops); the right algorithm instead
maps the equation onto two {\sc gemv}s,
with a cost of $O(n^2)$ flops.
The difference lays in
how the input operation is decomposed and mapped
onto available kernels. 
In more complex matrix equations, it is not uncommon to face 
dozens and dozens of alternative decompositions, all corresponding 
to viable, but not equally effective, algorithms. 
We will illustrate how unfruitful branches can be avoided 
by propagating knowledge, as the algorithm unfolds,  
from the input operands to intermediate results.

\begin{center}
\renewcommand{\lstlistingname}{Algorithm}
\begin{minipage}{0.40\linewidth}
	\begin{lstlisting}[caption={Matlab's algorithm for $x := Q^T L \, y$}, escapechar=!, label=alg:matlab-qly]
	$T := Q^T L$            !({\sc gemm})!
	$x := T y$            !({\sc gemv})!
\end{lstlisting}
\end{minipage}
\hspace*{2cm}
\begin{minipage}{0.40\linewidth}
	\begin{lstlisting}[caption={Alternative algorithm for $x := Q^T L \, y$}, escapechar=!, label=alg:compiler-qly]
	$t := L \, y$            !({\sc gemv})!
	$x := Q^T t$            !({\sc gemv})!
\end{lstlisting}
\end{minipage}
\end{center}

Challenging matrix equations appear in 
applications as diverse as 
machine learning, sensitivity analysis, and computational biology.
In most cases, one
has to solve not one instance of the problem, but thousands or even
billions of them: For example, the computation of mixed models in
the context of the genome-wide association study (GWAS), a popular study
in computational biology~\cite{10.1371/journal.pgen.1001256-short,Levy2009-short,Speliotes2010-short}, 
requires the solution of up to $10^{12}$ (trillions) instances of the equation
\[
 b := (X^T M^{-1} X)^{-1} X^T M^{-1} y,
 \quad {\rm where} \quad
 M = h^2 \Phi + (1 - h^2) I.
\]
Most
interestingly, these instances are not independent from one another,
suggesting that intermediate results could be saved and
reused; unfortunately, none of the current libraries allows this.

In order to overcome the deficiencies discussed so far,
we prototyped a linear algebra compiler, written in Mathematica, 
that takes as input a target equation annotated with properties, 
and returns as output a family of 
high-performance application-tailored algorithms.
Very much like a standard compiler takes a computer program and maps
it onto the instruction set provided by the processor, our approach is
to decompose the input equations into kernels provided by linear algebra libraries
such as BLAS and LAPACK~\cite{blas3,laug}.
As previously shown, the mapping is not unique, and the number of
alternatives may be very large. For this reason, our compiler carries
out a search within the space of possible algorithms, and yields the
most promising ones.  The search is guided by a number of heuristics
which, in conjunction with a mechanism for inferring properties, aim
at replicating and extending the thought-process of an expert in the
field.  Moreover, by means of dependency analyses, the compiler actively
seeks to avoid redundant computation, both within a single equation
and across sequences of them.  The combination of these techniques
produces remarkable speedups: We used the compiler to tackle operations
in genome analysis and sensitivity studies; in both cases, we attained
more than 100x speedups.

The heuristics used by our compiler 
are discussed in Section~\ref{sec:heuristics}, while 
the compiler's modular design is described in Section~\ref{sec:engine}.
A detailed example of the search process is presented in Section~\ref{sec:example}.
In Sections~\ref{sec:sequences} and~\ref{sec:results}, respectively, we cover sequences of problems and two sets of experiments.
Finally, conclusions and future work are given in Section~\ref{sec:conclusions}.

\paragraph{ {\bf Related work.} }

After more than a decade of extensive research on domain-specific
languages and libraries, automation has shown its benefits
in a broad variety of fields, including numerical linear 
algebra~\cite{atlas-sc98,Bientinesi:2005:SDD,Vuduc:2004uz,Baumgartner05synthesisof},
signal processing~\cite{FFTW05,Pueschel:05},
and differential equations~\cite{Logg:2010:DAF:1731022.1731030}.
The overarching theme is 
the attempt of exploiting domain knowledge 
to automatically generate highly efficient routines
while at the same time reducing coding and maintainability effort.

The approach adopted by most of these projects is that of exploring 
a parameter space through more or less sophisticated search mechanisms. 
ATLAS~\cite{atlas-sc98} and FFTW~\cite{FFTW05} provide
an optimized implementation of the BLAS, and an adaptive library for
Fourier Transforms, respectively. 
Both these libraries are based on the automatic performance tuning
of codelets; the search for the best codelet is steered empirically 
via actual execution and timing.
In SPIRAL~\cite{Pueschel:05}, a project targeting
high-performance implementations of transforms for 
digital signal processing, 
the search space comes from the combination of 
breakdown rules to decompose the transforms in a divide and conquer fashion, 
and parameterized rewrite rules to incorporate knowledge of the architecture.
In the field of linear algebra, 
DxTer~\cite{DxTer} starts from an LAPACK-like algorithm, and 
aims at replicating the process carried out by domain experts to
obtain efficient distributed-memory implementations.

In contrast to the aforementioned projects, our target consists of mathematical equations.
Our compiler makes use of heuristics---inspired by the thought-process
of human experts---to prune the search space
and tailor the algorithm to the specific application;
key to this process is the dynamic inference and exploitation of domain knowledge. 
Many of the optimizations used are the logic extension to matrix operands
of techniques used by traditional compilers on vectors and scalars~\cite{Aho:86}.

\section{Heuristics for the generation of algorithms}
\label{sec:heuristics}

Starting from a target equation, 
our compiler explores a subset of the space of possible algorithms, 
dynamically generating a ``tree of decompositions''.
For instance, Figure~\ref{fig:linsys-full} contains the complete tree 
generated for the solution of a linear system, 
when the coefficient matrix is symmetric positive definite (SPD).
The root node corresponds to the input equation, 
and every branch represents the mapping onto a building block;
in the example, 
the three branches are originated by three different factorizations of the matrix $A$. 
Once the process is over, 
the operations along the edges from the root to each leaf  
constitute a valid algorithm.
In practice, the tree is built in two phases, corresponding to 
the blue(dark) and green(light) nodes, respectively. 
In order to limit the size of the tree,
the compiler uses the heuristics described hereafter. 

\begin{figure}
\centering
	\centering
	\includegraphics[scale=0.6]{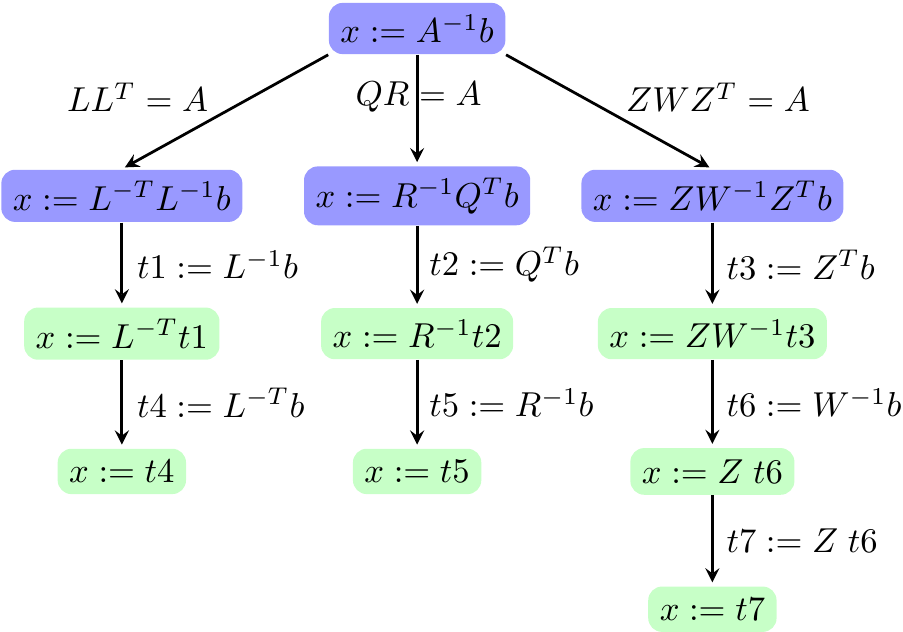}
	\label{fig:linsys-full}
	\caption{Full tree spawned by the compiler when processing the solution of a linear system of
		equations $x := A^{-1} b$, with an SPD coefficient matrix $A$, and a single right-hand
		side $b$.}
	\label{fig:linsys-full}
\end{figure}

\subsection{Dealing with the inverse operator}
\label{subsec:heuristics-one}

The inversion of matrices is a delicate operation.
There are only rare occurrences of problems in which one 
is interested in the actual matrix inverse;
most often, the operation appears in the context of linear systems, least squares problems,
or more complex expressions; in the majority of cases, the inversion 
can---and should---be avoided altogether. 
Because of this, our compiler splits the generation of algorithms 
in two phases, 
the first of which is solely devoted to the treatment of inverses;
the objective is to reduce the input equations to an expression in which 
the inverse is only applied to matrices in factored form,
i.e., triangular or diagonal (see blue subtree in Figure~\ref{fig:linsys-full}).
In the second phase, the resulting expression is 
mapped onto computational kernels (see green branches in Figure~\ref{fig:linsys-full}).

\begin{figure}
\centering
\includegraphics[scale=0.88]{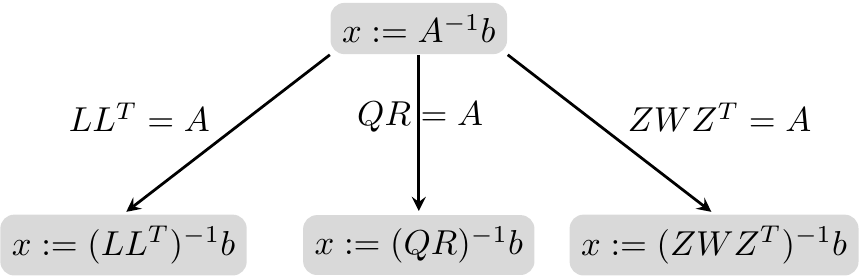}
\caption{Solution of an SPD linear system. 
In the first iteration, 
the compiler identifies three viable factorizations 
for the coefficient matrix $A$; this originates three
branches, corresponding to 
a Cholesky factorization (left), 
a QR factorization (middle), 
and an eigendecomposition (right).}
\label{fig:linsys-tree}
\end{figure}

This first phase takes as input the target equation, and
generates the subtree characterized by leaf nodes that require 
no further treatment of the inverses.
This is an iterative process in which the tree is constructed 
in a breadth-first fashion;
at each iteration, the current expression is inspected for inverse operators,
the innermost of which is then handled.
The inversion is applied to either a full matrix, such as $A^{-1}$, 
or to a non-simplifiable expression, e.g., $(A^T A)^{-1}$ with $A$ rectangular.
In the first case, the matrix is factored by means of one 
of the many matrix decompositions provided by LAPACK, 
but instead of exhaustively trying all possibilities,
the factorization is chosen according 
to the properties of the matrix.
For instance, if $A$ is 
a symmetric positive definite  matrix, 
viable  options are
the QR factorization ($QR = A$), 
the Cholesky factorization ($L L^T = A$), 
and the eigendecomposition ($Z W Z^T$); 
vice versa, the LU ($LU = A$), and LDL ($LDL^T = A$) 
factorizations are not considered.
As depicted in Figure~\ref{fig:linsys-tree}, 
the compiler constructs as many branches as factorizations,
while altering the initial expression. 
All the branches are subsequently  explored.

Limiting the search to a subset of all possible factorizations 
has two advantages:
On the one hand, non-promising algorithms are discarded
and the search space is pruned early on;
on the other hand, the algorithm is 
tailored to the specific properties of the application.
Table~\ref{tab:factorizations} contains 
the set of factorizations currently in use,
together with the matrix properties that enable them.

\begin{table}
\centering
\setlength\extrarowheight{2pt}
\renewcommand{\arraystretch}{1.0}
\begin{tabular}{l@{\hspace*{8mm}} l} \toprule
	{\bf Matrix Property} & {\bf Factorizations} \\\midrule 
	Symmetric & LDL, QR, Eigendecomposition \\
	SPD       & Cholesky, QR, Eigendecomposition \\
	Column Panel (FullRank) & QR \\
	Column Panel (RankDef) & SVD \\
	Row Panel (FullRank) & LQ \\
	Row Panel (RankDef)  & SVD \\
	General & LU, SVD \\\bottomrule
\end{tabular}
\caption{Factorizations currently used by the compiler, 
  and matrix properties that enable them.}
\label{tab:factorizations}
\end{table}

\begin{figure}
\centering
	{
		\centering
		\hspace{2mm}
		\includegraphics[scale=0.74]{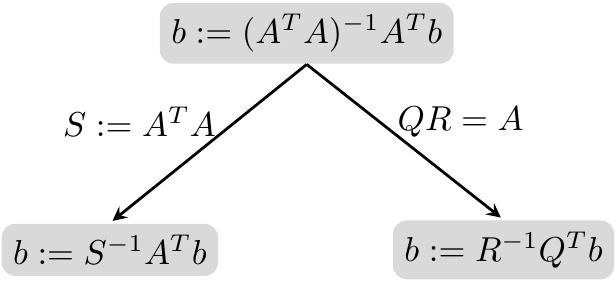}
		\hspace{2mm}
		\label{fig:ols_tree_fr}
	}
	\caption{Snippet of the tree spawned by the compiler when processing the ordinary least-squares equation
	$b := (A^T A)^{-1} A^T y$, where $A \in R^{m \times n} (m > n)$ is full rank.
        }
	\label{fig:ols} 
\end{figure}

We concentrate now on the case of  
an inverse operator applied to a non-simplifiable expression.
A characteristic example
is that of the normal equations, arising for instance as
part of the ordinary least-squares problem
\begin{equation}
	b := (A^T A)^{-1} A^T y,
  \label{eq:ols}
\end{equation}
where $A \in R^{m \times n}$ (with $m > n$) is full rank.
In this scenario, as depicted in Figure~\ref{fig:ols}, 
our compiler explores two alternative routes:
1) the multiplication of the expression $A^T A$, thus reducing it
to the inverse of a single SPD operand $S$; and
2) the decomposition of one of the matrices in the expression, 
in this case $A$, thus spawning a branch per 
suitable factorization.
As dictated by Table~\ref{tab:factorizations}, 
in Eq.~\eqref{eq:ols} $A$ is decomposed by means of a QR factorization.

The treatment of inverses continues
until the inverse operator is only applied to triangular or diagonal matrices.
For the example in Figure~\ref{fig:ols}, 
the left branch would be further processed 
by factoring the matrix $S$,
yielding three more nodes;
the right branch instead, 
since $R$ is a triangular matrix, 
is complete.

\subsection{Mapping onto kernels}
\label{subsec:mapping}

The goal of this second phase is 
to find efficient mappings from expressions
to kernels provided by numerical libraries,
i.e., BLAS and LAPACK.
The number of possible mappings grows exponentially with the number
of operators in the expression, therefore 
heuristics are necessary to constrain 
the amount of explored alternatives.
We discuss two examples of such heuristics.

\paragraph{ Common segments }
The objective is 
to reduce the complexity of the algorithm
by avoiding redundant computations;
common segments of the expression are identified,
thus allowing the reuse of intermediate results.
We emphasize that this is by no means a trivial optimization. 
In fact, 
even for the simplest cases,
sophisticated tools such as Matlab do not adopt it.
For instance, when computing the operation
$$ \alpha := x^T y x^Ty, $$
where $x$ and $y$ are vectors of size $n$, 
Matlab executes Algorithm~\ref{alg:matlab-comseg}; our
compiler recognizes that the expression $x^T y$ appears twice,
and instead generates Algorithm~\ref{alg:compiler-comseg}, 
which reduces the number of flops from $5n$ to $2n$.

\begin{center}
\renewcommand{\lstlistingname}{Algorithm}
\begin{minipage}{0.40\linewidth}
\begin{lstlisting}[caption=Matlab's computation for $x^T y x^T y $, escapechar=!, label=alg:matlab-comseg]
  t1 = x' * y
  t2 = t1 * x'
  alpha = t2 * y
\end{lstlisting}
\end{minipage}
\hspace*{2cm}
\begin{minipage}{0.40\linewidth}
\begin{lstlisting}[caption=Our compiler's code for $x^T y x^T y $, escapechar=!, label=alg:compiler-comseg]
  t1 = x' * y
  alpha = t1 * t1 !\vspace{4mm}!
\end{lstlisting}
\end{minipage}
\end{center}

More challenging is the case where one of the occurrences of
the common segment appears in transposed or inverted form.
As an example, let us consider the expression
$$ v := X^T L^{-1} L^{-T} X,$$
where both operands $X$ and $L$ are matrices, and $L$ is triangular.
In order to recognize that $L^{-T} X$ is the transpose of
$X^T L^{-1}$, and in general, to recognize that two segments are
the negation, inverse, or transpose of one another,
our compiler incorporates a large set of ground linear algebra
knowledge. This is covered in Section~\ref{sec:engine}.

\paragraph{ Prioritization }
In an attempt to minimize the cost of the generated algorithms, 
the kernels available to the compiler are classified according to
a precedence system. 
In Table~\ref{tab:precedence}, we give an example of a subset
of these kernels, sorted from high to low priority.
The precedences are driven by the dimensionality of the operands
in the kernels: The idea is to reduce the number of required flops
by keeping the dimensionality of the resulting operands as low as possible.
The first two kernels in the table reduce the dimensionality of
the output operand with respect to that of the input, while
the third kernel maintains it, and the fourth increases it. 
Finally, the inversion of a triangular matrix is given the lowest precedence: 
A matrix will only be inverted if no other option is available.

\begin{table}
	\centering
	\begin{tabular}{c @{\hspace{3mm}} l @{\hspace{3mm}} l @{\hspace{3mm}} c @{\hspace{3mm}} c @{\hspace{3mm}} c} \toprule
		{\bf \#} & {\bf Kernels} & {\bf Example} & {\bf Dim(op1)} & {\bf Dim(op2)} & {\bf Dim(out)} \\\midrule
	1 & inner product & $\alpha := x^T y$ & 1 & 1 & 0 \\
	2 & matrix-vector operations & $y := A x$, $b := L^{-1} x$ & 2 & 1 & 1 \\
	3 & matrix-matrix operations & $C := A B$, $B := L^{-1} A$ & 2 & 2 & 2 \\
	4 & outer product & $A := x y^T$ & 1 & 1 & 2 \\
	5 & inversion of a triangular matrix & $C := L^{-1}$ & -- & -- & -- \\\bottomrule
\end{tabular}
\caption{Example of the classification of kernels based on a system of precedences.
The kernels that reduce the dimensionality of the output operands with respect to the input ones are
given higher precedence. The inversion is only selected when no other option exists.}
\label{tab:precedence}
\end{table}

The benefits of the prioritization were already outlined in the Introduction
(Algorithms~\ref{alg:matlab-qly}~and~\ref{alg:compiler-qly}):
there, by favoring the matrix-vector over 
the matrix-matrix product, the complexity
was lowered by an order of magnitude. 
Here, we provide more examples.
Consider the operation
$$ \alpha := x^T z x^T y,$$
where $x$, $y$, and $z$ are vectors, and $\alpha$ is a scalar.
When inspecting the expression for kernels, the compiler finds 
two inner products ($x^T z$, and $x^T y$), and one outer product ($z x^T$).
While all three options lead to valid algorithms,
the inner products are favored,
producing, for instance, Algorithm~\ref{alg:inner}; 
the cost of this algorithm is $O(n)$ flops, instead of 
a cost of $O(n^2)$, had the compiler favored 
the outer product (Algorithm~\ref{alg:outer}).

\begin{center}
\renewcommand{\lstlistingname}{Algorithm}
\begin{minipage}{0.40\linewidth}
	\begin{lstlisting}[caption={Computation of $ \alpha := x^T z x^T y$, favoring inner products}, escapechar=!, label=alg:inner]
  t1 := x' * z
  t2 := x' * y
  alpha := t1 * t2
\end{lstlisting}
\end{minipage}
\hspace*{2cm}
\begin{minipage}{0.40\linewidth}
	\begin{lstlisting}[caption={Computation of $ \alpha := x^T z x^T y$, favoring outer products}, escapechar=!, label=alg:outer]
  T1 := z * x'
  t2 := x' * T1
  alpha := t2 * y
\end{lstlisting}
\end{minipage}
\end{center}

A third example is given by the linear system
\begin{equation}
	\beta := v^T L^{-1} L^{-T} u,
	\nonumber
\end{equation}
where $L$ is a square lower triangular matrix, and $v$ and $u$ are vectors.
The inspection for kernels yields the following matches: $v^T L^{-1}$, $L^{-1}$,
and $L^{-T} u$. However, the inversion of $L$ is avoided, unless no alternatives exist.
This is captured by the precedences listed in Table~\ref{tab:precedence}, which give priority to the solution
of linear systems over the inversion of matrices. Therefore, the second
option ($L^{-1}$) is dismissed, and the compiler only explores the branches
spawned by the first and third kernels. While the inversion of $L$ would lead to
a cubic algorithm (Algorithm~\ref{alg:matlab-inv}), the ones generated 
(e.g., Algorithm~\ref{alg:compiler-trsv}) have a quadratic cost.

\begin{center}
\renewcommand{\lstlistingname}{Algorithm}
\begin{minipage}{0.40\linewidth}
	\begin{lstlisting}[caption={Computation of $\beta := v^T L^{-1} L^{-T} u$,
	favoring the inversion of matrices}, escapechar=!, label=alg:matlab-inv]
	T1 := inv(L)
	t2 := v' * T1
	t3 := t2 * T1'
	beta := t3 * u
\end{lstlisting}
\end{minipage}
\hspace*{2cm}
\begin{minipage}{0.40\linewidth}
	\begin{lstlisting}[caption={Computation of $\beta := v^T L^{-1} L^{-T} u$,
	favoring the solution of triangular systems}, escapechar=!, label=alg:compiler-trsv]
	t1 := v' / L
	t2 := L' \ u
	beta := t1 * t2 !\vspace{3.8mm}!
\end{lstlisting}
\end{minipage}
\end{center}

Notice that if implemented naively, the rules discussed so far 
may lead to an infinite process: 
For instance, a matrix could be factored and built again, 
as in 
$(A^T A)^{-1} \xrightarrow{QR = A} ((QR)^T QR)^{-1} \xrightarrow{A := QR} (A^T A)^{-1}$; 
also, a matrix could be factored indefinitely, as in 
$A \xrightarrow{Q_1 R_1 = A} Q_1 R_1
\xrightarrow{Q_2 R_2 = Q_1} Q_2 R_2 R_1 
\xrightarrow{\dots} \dots 
\xrightarrow{Q_i R_i = Q_{i-1}} Q_i R_i \dots R_2 R_1 $.
To avoid such situations, our compiler incorporates a mechanism to measure and guarantee progress.

\section{Compiler's engine}
\label{sec:engine}

The availability of knowledge 
is crucial for a successful application of the heuristics.
Equally important is the capability of algebraically manipulating
expressions with the objective of simplifying them 
or finding common segments.
Here, we detail the different modules that constitute
the compiler's engine, and how these modules enable:
1) the algebraic manipulation of expressions,
2) the mapping onto building blocks, and
3) the management of both input and inferred knowledge.

\subsection{Matrix algebra}
\label{subsec:algebra}

The {\em Matrix algebra} module deals with the algebraic
manipulation of expressions.
It incorporates a considerable amount of knowledge regarding
properties of the operators, such commutativity 
and distributivity, and linear algebra equalities,
such as
``the inverse of an orthogonal matrix equals its transpose''.
This knowledge is encoded as 
an extensive list of {\em rewrite rules} that allow the compiler to
rearrange expressions, simplify them, and find subexpressions that are the 
inverse, transpose, etc, of one another. 

A rewrite rule consists of a left-hand and a right-hand side.
The left-hand side contains a pattern, possibly restricted via constraints to be satisfied by the operands;
the right-hand side specifies how the pattern, if matched, should be replaced.
For instance, the rule
\begin{center}
{\tt inv[Q\_] /; isOrthogonal[Q] -> trans[Q]}
\end{center}
reads as follows: The inverse of a matrix $Q$, provided that $Q$ is orthogonal,
may be replaced with the transpose of $Q$.
Box~\ref{box:simplify-rules}
includes more examples of rewrite rules.
\begin{mybox}
	\vspace{1em}
	\centering
	\begin{minipage}{0.90\textwidth}
  {\tt
    trans[times[A\_, B\_]] -> times[trans[B], trans[A]]; \\
    times[trans[Q\_], Q\_] /; isOrthogonalQ[Q] -> Identity; \\
	times[A\_, Identity] /; Not[ isScalarQ[A] ] -> A; \\
    inv[times[A\_, B\_]] /; isSquareQ[A] \&\& isSquareQ[B] -> times[inv[B], inv[A]]; \\
    times[inv[A\_], A\_] -> Identity; \\
  }
  \end{minipage}
  \caption{Rewrite rules used for the transformations shown in Box~\ref{box:simplification}.}
  \label{box:simplify-rules}
\end{mybox}

The example in Box~\ref{box:simplification} gives an idea of
how the compiler is capable of 
eliminating unnecessary calculations
by means of algebraic transformations.
As dictated by the heuristics presented in Section~\ref{sec:heuristics},
one way of handling the initial expression $(X^T X)^{-1} X^T L^{-1} y$ 
is through a QR factorization of the matrix $X$:
The symbol $X$ is replaced by $Q R$---line 2---(where 
$Q$ and $R$ are an orthogonal and an upper triangular matrix, respectively),
and a series of transformations are triggered. 
First, the transposition is distributed over the product---line 3---;
next, due to the orthogonality of $Q$, the product 
$Q^T Q$ is removed as it equals the identity---line 4---.
Since $R$ is square, the inverse may be distributed 
over the product $R^T R$ resulting in $R^{-1} R^{-T}$---line 5---. 
Another simplification rule establishes that the product 
of a square matrix with its inverse equals the identity; because of this, 
the $R^{-T} R^T$ is removed---line 6---. 
After all these algebraic steps, 
the expression $((Q R)^T Q R)^{-1} (Q R)^T L^{-1} y$
simplifies to $R^{-1} Q^T L^{-1} y$. 
Box~\ref{box:simplify-rules} contains the necessary 
set of rewrite rules for this manipulation.

\begin{mybox}
\begin{align}
	1) \quad b &:= (X^T X)^{-1} X^T L^{-1} y; \nonumber \\
	2) \quad b &:= ((Q R)^T Q R)^{-1} (Q R)^T L^{-1} y; \nonumber \\
	3) \quad b &:= (R^T Q^T Q R)^{-1} R^T Q^T L^{-1} y; \nonumber \\
	4) \quad b &:= (R^T R)^{-1} R^T Q^T L^{-1} y; \nonumber \\
	5) \quad b &:= R^{-1} R^{-T} R^T Q^T L^{-1} y; \nonumber \\
	6) \quad b &:= R^{-1} Q^T L^{-1} y. \nonumber
\end{align}
\caption{Example of expression simplification carried out by the compiler.}
\label{box:simplification}
\end{mybox}

Such rewrite rules are algebraic identities, i.e., 
they may be applied in both directions. 
For instance, the expression $(A B)^T$ may be rewritten as $B^T A^T$, and vice versa,
leading to multiple equivalent representations for the same expression. 
Since this fact complicates the manipulation and identification 
of building blocks, one may be tempted to use rules as
``always distributing the transpose over the product''
for reducing every expression to a canonical form.
Unfortunately, there exists no a ``best'' representation for expressions. 
Indeed, imposing a canonical form would lower the effectiveness of the compiler.

A prototypical example is given by the distribution of the product
over the addition: 
$(A + B)C$ may be transformed into $A C + B C$ and vice versa, 
but neither representation is superior in all scenarios.
Consider, for instance, the expression 
$ \alpha x x^T + \beta y x^T + \beta x y^T$,
where $\alpha$ and $\beta$ are scalars, and $x$ and $y$ are vectors.
In this format, 
it is straightforward to realize that the expression is 
symmetric---the first term is symmetric, and the second
and third are one the transpose of the other---;
if instead $x^T$ is factored out as in 
$ (\alpha x + \beta y) x^T + \beta x y^T $,
the symmetry is not visible, 
and redundant computation would be performed.
This is an example in which the distribution of the product over the addition 
seems to be the choice to favor.

On the contrary, let us consider the expression in Box~\ref{box:factor-out}:
$(Z W Z^T + Z Z^T)^{-1},$ where $Z$ is square and orthogonal, and $W$ is diagonal. 
Factoring $Z$ and $Z^T$ out---$(Z (W + I) Z^T)^{-1}$, where $I$ is the identity matrix---is 
an indispensable first step towards a simplification of the expression.
Next, since all matrices are square, 
the inverse may be distributed over the product, 
and the orthogonality of $Z$ allows the rewriting of its inverse as its transpose, 
resulting in $Z (W + I)^{-1} Z^T$. 
This transformation---absolutely crucial in practical cases---is only possible 
thanks to the initial factoring;
hence, this is a contrasting example 
in which the distribution of the product is not the best option.
In light of this dichotomy, 
our compiler always operates with multiple alternative representations.

\begin{mybox}
\begin{align}
M &:= (Z W Z^T + Z Z^T)^{-1}; \nonumber \\
M &:= (Z (W + I) Z^T)^{-1}; \nonumber \\
M &:= Z^{-T} (W + I)^{-1} Z^{-1}; \nonumber \\
M &:= Z (W + I)^{-1} Z^T; \nonumber
\end{align}
\caption{Example of expression manipulation carried out by the compiler.}
\label{box:factor-out}
\end{mybox}

\subsection{Interface to building blocks}

We have claimed repeatedly that the goal of the compiler is to decompose 
the target equation in terms of building blocks that 
can be directly mapped to library invocations;
it remains to be discussed what are the available building blocks. 
The exact list is configurable,
and is provided to the compiler via the {\em Interface to building blocks} module.
This module contains 
a list of patterns associated to the corresponding computational kernels.
As of now, this list includes a subset of the operations provided by 
BLAS and LAPACK, e.g.,
matrix factorizations, matrix products, and the solution of linear systems;
a sample is given in Box~\ref{box:bblocks-sample}.

\begin{mybox}
	\centering
	\begin{minipage}{0.80\textwidth}
  {\sc Factorizations:} \\[1mm]
  {\tt 
    \hspace*{2mm} equal[ times[ L\_, U\_], A\_ ] /; isLowerQ[L] \&\& isUpperQ[U] \\
    \hspace*{2mm} equal[ times[ L\_, trans[L\_]], A\_ ] /; isLowerQ[L] \&\& isSPDQ[A] \\
    \hspace*{2mm} equal[ times[ Q\_, R\_], A\_ ] /; isOrthogonalQ[Q] \&\& isUpperQ[R] \\[2mm]
  }
  {\sc Matrix products:} \\[1mm]
  {\tt 
    \hspace*{2mm} plus[ times[ alpha\_, A\_, B\_], times[beta\_, C\_] ] \\
    \hspace*{2mm} plus[ times[ alpha\_, trans[A\_], B\_], times[beta\_, C\_] ] \\
    \hspace*{2mm} plus[ times[ alpha\_, trans[A\_], A\_], times[beta\_, C\_] ] \\
    \hspace*{2mm} times[ A\_, trans[A\_] ] /; isTriangularQ[A] \\[2mm]
  }
  {\sc Linear systems:} \\[1mm]
  {\tt
    \hspace*{2mm} plus[ times[ inv[A\_], B\_ ] ] /; isTriangularQ[A] \&\& isMatrixQ[B] \\
    \hspace*{2mm} plus[ times[ inv[A\_], b\_ ] ] /; isTriangularQ[A] \&\& isVectorQ[b] 
  }
  \end{minipage}
  \caption{A snippet of the interface to available building blocks.
  }
  \label{box:bblocks-sample}
\end{mybox}

The compiler is by no means limited to this set of operations. 
Should an additional or a different set of building blocks be available, 
say RECSY~\cite{RECSY1,RECSY2} 
or an extension of the BLAS library~\cite{2002:USB:567806.567807}, 
this can be made accessible to the compiler with only minimal effort,
by including in this module the corresponding patterns. 
For instance, in order to add support for the operation $w := \alpha x + \beta y$, 
as proposed in an extension of the BLAS library~\cite{2002:USB:567806.567807}, 
we only need to incorporate the pattern
\begin{verbatim}
     plus[ times[ alpha_, x_ ], times[ beta_, y_ ] ] /;
                  isVectorQ[x,y] && isScalarQ[alpha, beta];
\end{verbatim}
The compiler is then ready to make use of this building block in the generation of algorithms.

\subsection{Inference of properties}

As discussed throughout the paper,
properties play a central role in the search for efficient algorithms;
the more knowledge is available,
the more opportunities arise for further optimizations.
A distinguishing feature of our compiler is the propagation
of properties: We developed an engine for
inferring properties of expressions from those of the individual operands.
Thanks to this engine, the initial knowledge (from the input equation) is augmented dynamically.

This mechanism is activated every time a mapping takes place:
1) when mapping onto factorizations, properties
are propagated from the input matrix to its factors;
2) when mapping onto kernels, properties are propagated
from the segment to the output quantity.
The gained knowledge on the intermediate operands is then used
by the compiler for tailoring the algorithms.
Boxes~\ref{box:inf-fact}~and~\ref{box:inf-kernels} provide
examples of inference of knowledge in factorizations and kernels,
respectively.

\begin{mybox}
	\centering
	\begin{minipage}{.6\textwidth}
		{\sc eigendecomposition} ($Z W Z^T = A$): \\
		\begin{tabular}{@{\hspace{5mm}}l @{\hspace{5mm}}l}
			Input  & $A$: matrix, symmetric \\
			Output & $Z$: matrix, square, orthogonal \\
				   & $W$: matrix, square, diagonal \\
		\end{tabular}

		\vspace{3mm}

		{\sc qr} ($Q R = A$): \\
		\begin{tabular}{@{\hspace{5mm}}l @{\hspace{5mm}}l }
			Input  & $A$: matrix, column-panel, full rank \\
			Output & $Q$: matrix, orthogonal, column-panel, full rank \\
				   & $R$: matrix, square, upper triangular, full rank
		\end{tabular}
	\end{minipage}
	\caption{Inference of properties for two representative factorizations. }
	\label{box:inf-fact}
\end{mybox}

\begin{mybox}
	\centering
	\begin{minipage}{.6\textwidth}
		$W = L^{-1} X$: \\
		\begin{tabular}{@{\hspace{5mm}}l @{\hspace{5mm}}l }
			Input  & $L$: matrix, square, full rank \\
				   & $X$: matrix, column-panel, full rank \\
			Output & $W$: matrix, column-panel, full rank \\
		\end{tabular}

		\vspace{3mm}

		$S = W^T W$: \\
		\begin{tabular}{@{\hspace{5mm}}l @{\hspace{5mm}}l }
			Input  & $W$: matrix, column-panel, full rank \\
			Output & $S$: matrix, square, SPD \\
		\end{tabular}
	\end{minipage}
	\caption{Inference of properties for two mappings onto kernels. }
	\label{box:inf-kernels}
\end{mybox}

It is important to notice that the inference of rules and the mapping onto kernels are
completely independent actions. 
For instance, in the absence of the second rule in Box~\ref{box:inf-kernels},
the compiler would still be able to match a product of the form $A^T A$ (provided the pattern is
included in the {\em Interface to building blocks} module);
however, if $A$ is a full rank, column panel matrix, the compiler would not be able to
infer, and then exploit, the positive definiteness of $S$.

We regard the inference engine as a growing database of linear algebra knowledge.
In its current form, the database is populated with a sample of rules and theorems,
but the flexible design of the module allows it to be easily extended
with new inference rules.

\section{A detailed example}
\label{sec:example}

We use a challenging operation arising in computational 
biology---the genome-wide association study (GWAS)---to 
illustrate the potential of the compiler's engine and heuristics.
As part of GWAS, one has to solve the equation
\begin{equation}
\label{eq:probDef}
\left\{ 
{\begin{aligned}
	b_{ij} & := (X_i^T M_j^{-1} X_i)^{-1} X_i^T M_j^{-1} y_j \\
	M_j   & = h_j \Phi + (1 - h_j) I
\end{aligned}}
\right.
\;
\quad \text{ with }
{\begin{aligned}
 & 1 \le i \le m \\
 & 1 \le j \le t,
\end{aligned}}
\end{equation}
where $X_i$, $M_j$, and $y_j$ are known quantities, and $b_{ij}$ is sought after.
The size and properties of the operands are as follows:
$b_{ij} \in \mathcal{R}^{p}$, 
$X_i \in \mathcal{R}^{n \times p}$ is a full rank column panel ($n > p$), 
$M_j \in \mathcal{R}^{n \times n}$ is symmetric positive definite,
$y_j \in \mathcal{R}^{n}$, 
$\Phi \in \mathcal{R}^{n \times n}$,
$h_j \in \mathcal{R}$, and
$I$ is the identity matrix.
Box~\ref{box:input} contains the input representation of Eq.~\eqref{eq:probDef}. 

\begin{mybox}[!h]
\vspace{3mm}
\begin{verbatim}
    equation = {
        equal[ b,
           times[ inv[times[trans[X], inv[M], X]], trans[X], inv[M], y ]
        ],
        equal[ M, 
            plus[ times[h, Phi], times[plus[1, minus[h]], id] ]
        ]
    };
      
    operandProperties = {
        {X,   {``Input'',  ``Matrix'', ``ColumnPanel'', ``FullRank''} },
        {y,   {``Input'',  ``Vector'' } },
        {Phi, {``Input'',  ``Matrix'', ``Symmetric''} },
        {h,   {``Input'',  ``Scalar'' } },
        {M,   {``Input'',  ``Matrix'', ``SPD''} },
        {b,   {``Output'', ``Vector'' } }
    };

    dependencies = {{X, {i}}, {y, {j}}, {Phi, {}}, {h , {j}}, {M, {j}}, {b, {i,j}}};
\end{verbatim}
\caption{Mathematica description of GWAS as input to the compiler.}
\label{box:input}
\end{mybox}

\begin{figure}
\centering
\includegraphics[scale=0.88]{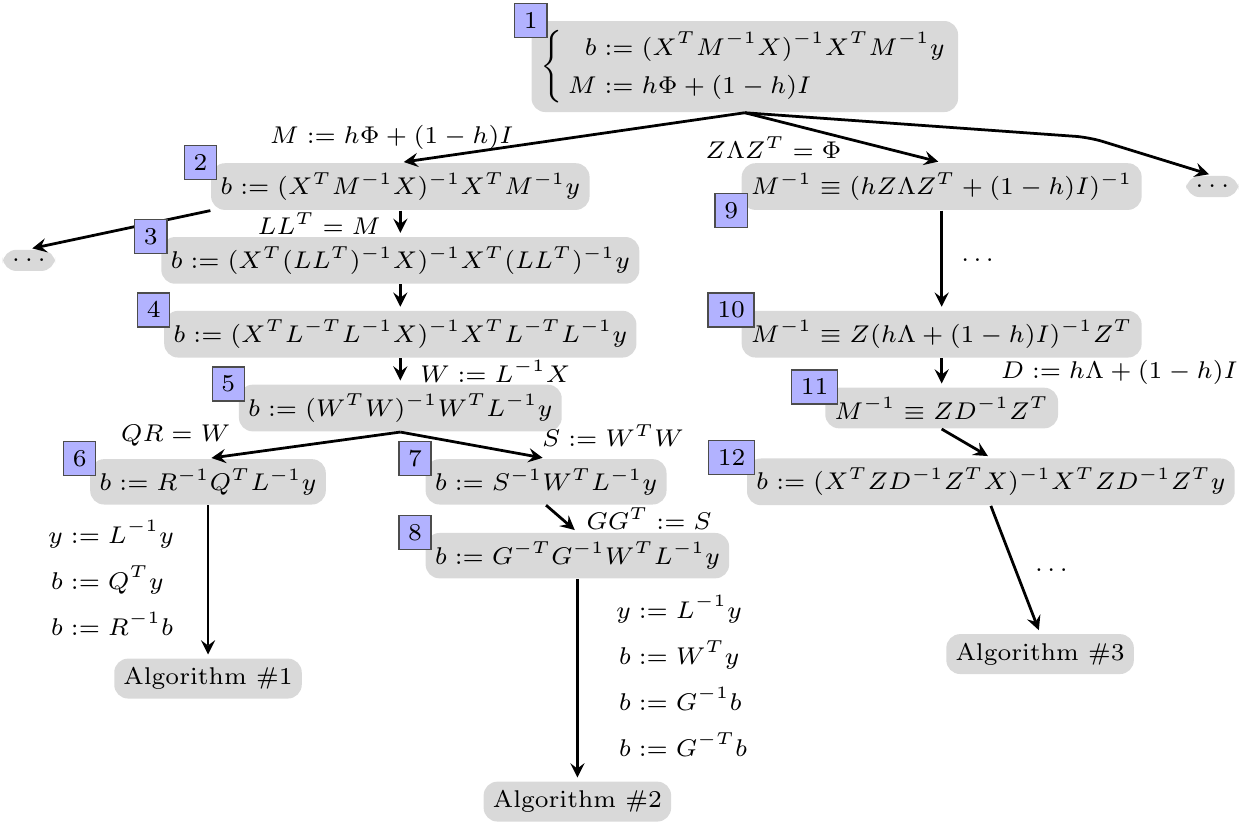}
\caption{Snippet of the tree spawned when creating algorithms
for the computation of GWAS.}
\label{fig:gwas-tree}
\end{figure}

Due to the complexity of GWAS, a large
number of alternatives is generated. For the sake of this discussion, we focus on the 
solution of a single instance of Eq.~\eqref{eq:probDef}, as if both $m$ and $t$ were 1.
In Figure~\ref{fig:gwas-tree}, we provide a snippet of the tree spawned by the compiler
while constructing algorithms. Among the dozens of different branches,
we describe three representative ones. 

At the root node,
 the compiler starts by dealing with the innermost inverse, 
$M^{-1}$, and equivalently, $ (h_j \Phi + (1 - h_j) I)^{-1} $. 
As explained in Section~\ref{sec:heuristics}, 
the options are either reduce the expression 
to a single operand ($M$, which is known to be SPD),
or factor one of the matrices in the expression, in this case $\Phi$. 
The former choice leads directly to Node 2 
(modulo the order in which addition and scaling are performed), 
while the latter opens up a number of branches, corresponding to 
all the admissible factorizations of $\Phi$;
the middle branch in Figure~\ref{fig:gwas-tree}
follows the eigendecomposition of $\Phi$.
One might argue that
based on the available knowledge ($M$ is SPD),
the compiler should decide against the eigendecomposition,
since a Cholesky factorization is about ten times as fast.
In actuality, although the decomposition is suboptimal for the solution of one single instance,
in the general case (Eq.~\eqref{eq:probDef})
it leads to the fastest algorithms of all.
\footnote{
This is because the eigendecomposition can be reused across
the entire two-dimensional sequence, 
while the Cholesky factorization cannot.
}

Let us concentrate on the subtree rooted at Node 2. 
The input equation was reduced to $b := (X^T M^{-1} X)^{-1} X^T M^{-1} y$; 
again, the compiler looks for the innermost inverse, $M^{-1}$, and
spawns a branch per factorization allowed for SPD matrices: 
QR, Cholesky, and eigendecomposition (Table~\ref{tab:factorizations});
here, we only describe the Cholesky factorization ($L L^T = M$),
which generates Node 3:
The equation becomes $b := (X^T (L L^T)^{-1} X)^{-1} X^T (L L^T)^{-1} y$, and  
the inference engine asserts a number of properties for $L$: 
square, lower triangular, and full rank.
The innermost inverse now is $(L L^T)^{-1}$;
since $L$ is square, rewrite rules allow the distribution of the 
inverse over the product $L L^T$, resulting in Node 4.

Once more, the compiler looks at the innermost inverse operators: 
In this case, they all are applied to triangular matrices, i.e., 
they do not require further treatment.
Therefore the focus shifts on the expression $(X^T L^{-T} L^{-1} X)^{-1}$;
$L$ is already in factored form, 
while according to a progress measure,
the factorizations of $X$ are not useful;
hence the compiler resorts to mappings onto kernels.
Matching the expression against the list of available kernels yields two
segments: $L^{-1}$ and $L^{-1} X$. 
The latter has higher priority,
so it is exposed ($W := L^{-1} X$), 
every occurrence is replaced with $W$ (generating Node 5),
and it is established that $W$ is a full rank, column panel (Box~\ref{box:inf-kernels}).

Similarly to the example depicted in Figure~\ref{fig:ols}, 
the inspection of Node 5
causes two branches to be constructed:
In the right one, the compiler multiplies out $S := W^T W$,
producing the $SPD$ matrix $S$ (Node 7).
In the left one, in accordance to the properties of $W$, 
the matrix is factored via a QR factorization;
after replacing $W$ with the product $Q R$, the simplifications
exposed in Box~\ref{box:simplification} are carried out, 
resulting in Node 6.
At this point, all inverses are processed, as 
the remaining ones are only applied to triangular matrices. 
For this node, the first phase (as described in Section~\ref{subsec:heuristics-one}) is completed, thus
the remaining expression is now to be mapped onto available building blocks. 
The compiler identifies the following kernels: 
$R^{-1}$, $R^{-1} Q^T$, $Q^T L^{-1}$, $L^{-1}$, and $L^{-1} y$.
The first four are either matrix inversions or matrix-matrix operations, 
while the last one corresponds to a matrix-vector operation. 
Based on the list of priorities (Table~\ref{tab:precedence}), 
the matrix-vector operation $L^{-1} y$ is chosen. 
The same reasoning is applied subsequently, 
leading to the sequence of operations 
$y' := L^{-1} y$,
$b := Q^T y'$, 
and $b := R^{-1} b$. 
A similar discussion leads from Node 7 to Algorithm \#2.

Finally, we focus on the subtree rooted at Node 9. After the eigendecomposition
of $\Phi$, the innermost inverse is given by $M^{-1} \equiv (h Z \Lambda Z^T + (1-h)I)^{-1}$.
Similar to the reasoning previously illustrated in Box~\ref{box:factor-out}, 
the compiler carries out a number of algebraic transformations that lead to the simplified
expression $M^{-1} \equiv Z (h \Lambda + (1-h)I)^{-1} Z^T$ (Node 10). 
Here, the innermost inverse is applied to a diagonal object ($\Lambda$ is diagonal and $h$ a scalar);
no more factorizations are needed, and $D := h \Lambda + (1-h)I$ is exposed (Node 11). 
The inverse of $M$ is then replaced in $b := (X^T M^{-1} X)^{-1} X^T M^{-1} y$, resulting in Node 12.
The subsequent steps develop similarly to the case of Node 4, generating Algorithm \#3.

Once the search is completed, the algorithms
are built by assembling the operations that label each edge along the path
from the root node to each of the leafs. 
The three algorithms are provided in 
Algorithms~\ref{alg:alg-qr},~\ref{alg:alg-chol}~and~\ref{alg:alg-eigen}.

\begin{center}
\renewcommand{\lstlistingname}{Algorithm}
\begin{minipage}{0.30\linewidth}
\begin{lstlisting}[caption=\normalsize \sc qr-gwas, escapechar=!, label=alg:alg-qr]
  $M := h\Phi + (1-h)I$
  $L L^T$ = $M$
  $W := L^{-1} X$
  $Q R = W$
  $y := L^{-1} y$
  $b := Q^T y$
  $b := R^{-1} b$ !\vspace{12.5mm}!
\end{lstlisting}
\end{minipage}
\hfill
\begin{minipage}{0.30\linewidth}
\begin{lstlisting}[caption={\normalsize \sc chol-gwas}, escapechar=!, label=alg:alg-chol]
  $M := h\Phi + (1-h)I$
  $L L^T = M$
  $W := L^{-1} X$
  $S := W^T W$
  $G G^T = S$
  $y := L^{-1} y$
  $b := W^T y$
  $b := G^{-1} b$
  $b := G^{-T} b$ !\vspace{3.5mm}!
\end{lstlisting}
\end{minipage}
\hfill
\begin{minipage}{0.30\linewidth}
\begin{lstlisting}[caption={\normalsize \sc eig-gwas}, escapechar=!,label=alg:alg-eigen]
  $Z \Lambda Z^T$ = $\Phi$
  $D := h \Lambda + (1 - h) I$
  $K := X^T Z$
  $V := K D^{-1}$
  $A := V K^T$
  $Q R = A$
  $y := Z^T y$
  $b := V y$
  $b := Q^T b$
  $b := R^{-1} b$
\end{lstlisting}
\end{minipage}
\end{center}

\section{Extensions: Sequences of related problems} 
\label{sec:sequences}

It is not uncommon that scientific and engineering applications require
the solution of not a single instance of a problem, but a sequence of them.
Typically, libraries and languages follow a black-box approach, 
i.e., they provide a routine to solve one instance, 
and this is then used repeatedly for the entire sequence. 
While this approach is acceptable for problems that are completely 
independent from one another, 
its rigidity leads to a suboptimal strategy when the problems are related, 
and intermediate results may be reused.
To overcome this limitation, our compiler breaks the black-box approach by
1) exposing the computation within the single-instance algorithm, 
2) performing an analysis of data dependencies, and 
3) rearranging the operations so that redundant computations are avoided. 

To illustrate the process, we look at the analysis of sensitivities~\cite{0021725}.
Here, one is interested in measuring how much
a simulation model is influenced by a set of parameters.
For each of the parameters, one instance of 
a problem similar to the one under scrutiny has to be solved.
We choose an equation arising as part of the analysis of an SPD linear system:
\begin{equation}
	x_i := C^{-1}(b_i - A_i y), \quad \text{with} \;\; 1 < i < p,
	\label{eq:sens}
\end{equation}
where $C \in R^{n \times n}$ is SPD, $A \in R^{n \times n}$ is symmetric,
and $x$, $b$, and $y \in R^n$.
The quantities
$C$, $b$, $A$, and $y$ are known, and $x$ is to be computed.
The input to the compiler and the index dependencies are provided in Box~\ref{box:input-sens}.
\begin{mybox}[!h]
\vspace{3mm}
\begin{verbatim}
         equation = {
            equal[ x, times[ inv[C], plus[ b, minus[times[A, y]] ] ] ]
         };

         operandProperties = {
            {C,   {``Input'',  ``Matrix'', ``Symmetric''} },
            {A,   {``Input'',  ``Matrix'', ``SPD''} },
            {b,   {``Input'',  ``Vector'' } },
            {y,   {``Input'',  ``Vector'' } },
            {x,   {``Output'', ``Vector'' } }
         };

         dependencies = { {x, {i}}, {b, {i}}, {A, {i}}, {C, {}}, {y, {}} };
\end{verbatim}
\caption{Compiler's input corresponding to the sensitivities Eq.~\eqref{eq:sens}.}
\label{box:input-sens}
\end{mybox}

The generation of algorithms for sequences of problems is divided in two steps.
First, the compiler creates a family of algorithms for a single instance, 
$x := C^{-1}(b-Ay)$,
via the techniques described in the previous section;
for example, Algorithm~\ref{alg:one-sensitivity} is produced.
Then, each of the algorithms is customized for the solution of the entire sequence.
A description of the latter step follows.

\begin{center}
\renewcommand{\lstlistingname}{Algorithm}
\begin{minipage}{0.35\linewidth}
\begin{lstlisting}[caption=Solution of a single sensitivity problem, escapechar=!, label=alg:one-sensitivity]
  $L L^T = C$
  $w := b - A y$
  $x := L^{-1} w$
  $x := L^{-T} x$
\end{lstlisting}
\end{minipage}
\end{center}

The single-instance algorithm is wrapped with as many loops as different indices;
in the case of Algorithm~\ref{alg:one-sensitivity}, 
this is a single loop along the $i$ dimension. 
Next, the compiler proceeds to identify operations that are 
loop-invariant, i.e., that do not depend on the indices;
then, it applies the {\em code motion} optimization, moving such
operations to the preheader of the loop.

In details,  invariant operations are identified analyzing the
dependencies between operands and loop indices: 
The compiler labels each operand
according to the input description.
The subscripts are then propagated with
a single-pass, from top to bottom, through the algorithm:
For each operation, 
the union of the indices appearing in the right-hand side
is attached to the operand(s) on the left-hand side, 
and to all their occurrences thereafter.
Algorithm~\ref{alg:one-sensitivity} before and after the labeling is presented here below.
$$
\begin{aligned}
	& L L^T = C  \\
	& w := b_i - A_i y  \\
	& x := L^{-1} w  \\
	& x := L^{-T} x 
\end{aligned}
\quad\quad
\longrightarrow
\quad\quad
\begin{aligned}
	& L L^T = C   \\
	& w_i := b_i - A_i y   \\
	& x_i := L^{-1} w_i   \\
	& x_i := L^{-T} x_i.
\end{aligned}
$$
At this point, 
any operation whose left-hand side does not include any subscript is 
invariant and will therefore be moved prior to the loop. This means that
the computation is performed once and 
reused in all successive loop iterations.
See  Algorithm~\ref{alg:many-sensitivities}
for the final rearrangement.

\begin{center}
\renewcommand{\lstlistingname}{Algorithm}
\begin{minipage}{0.35\linewidth}
\begin{lstlisting}[caption=gSPD - Solution of the sequence of sensitivities, escapechar=!, label=alg:many-sensitivities]
  $L L^T = C$
  for $i$ in 1..$p$
    $w_i := b_i - A_i y$
    $x_i := L^{-1} w_i$
    $x_i := L^{-T} x_i$
\end{lstlisting}
\end{minipage}
\end{center}

Let us quantify the gain obtained from the tailoring for sequences.
A traditional library would include a routine to solve one instance of Eq.~\eqref{eq:sens},
namely Algorithm~\ref{alg:one-sensitivity},
with a computational cost of $O(n^3)$; 
this routine would be used for each problem
in the sequence, for a total cost of $O(p n^3)$.
Instead, gSPD (Algorithm~\ref{alg:many-sensitivities})
tackles the sequence in its entirety, 
for a cost of $O(n^3) + O(p n^2)$.
As exposed by the experimental results (Section~\ref{sec:results}),
such an improvement in the computational complexity
leads to impressive speedups.

\section{Performance Results} \label{sec:results}

We present now performance results for the two applications discussed
in this paper: genome-wide association studies (Eq.~\eqref{eq:probDef}) and analysis
of sensitivities for an SPD linear system (Eq.~\eqref{eq:sens}). In both cases, we compare
the performance of the routine implementing our best algorithm
with broadly-used tools in the respective fields.
The results attest the potential of our compiler.

The experiments were performed on an SMP system consisting of 4 Intel Xeon E7-4850
multi-core processors. Each processor comprises 10 cores operating at 2 GHz.
The system is equipped with 512 GB of RAM.
The routines were compiled using the GNU C (version 4.4.5) and Fortran (version 4.4.6) 
compilers, and linked to the Intel MKL library, version 12.1.

\begin{figure}
	\begin{minipage}[t]{.45\textwidth}
\centering
\includegraphics[scale=0.40]{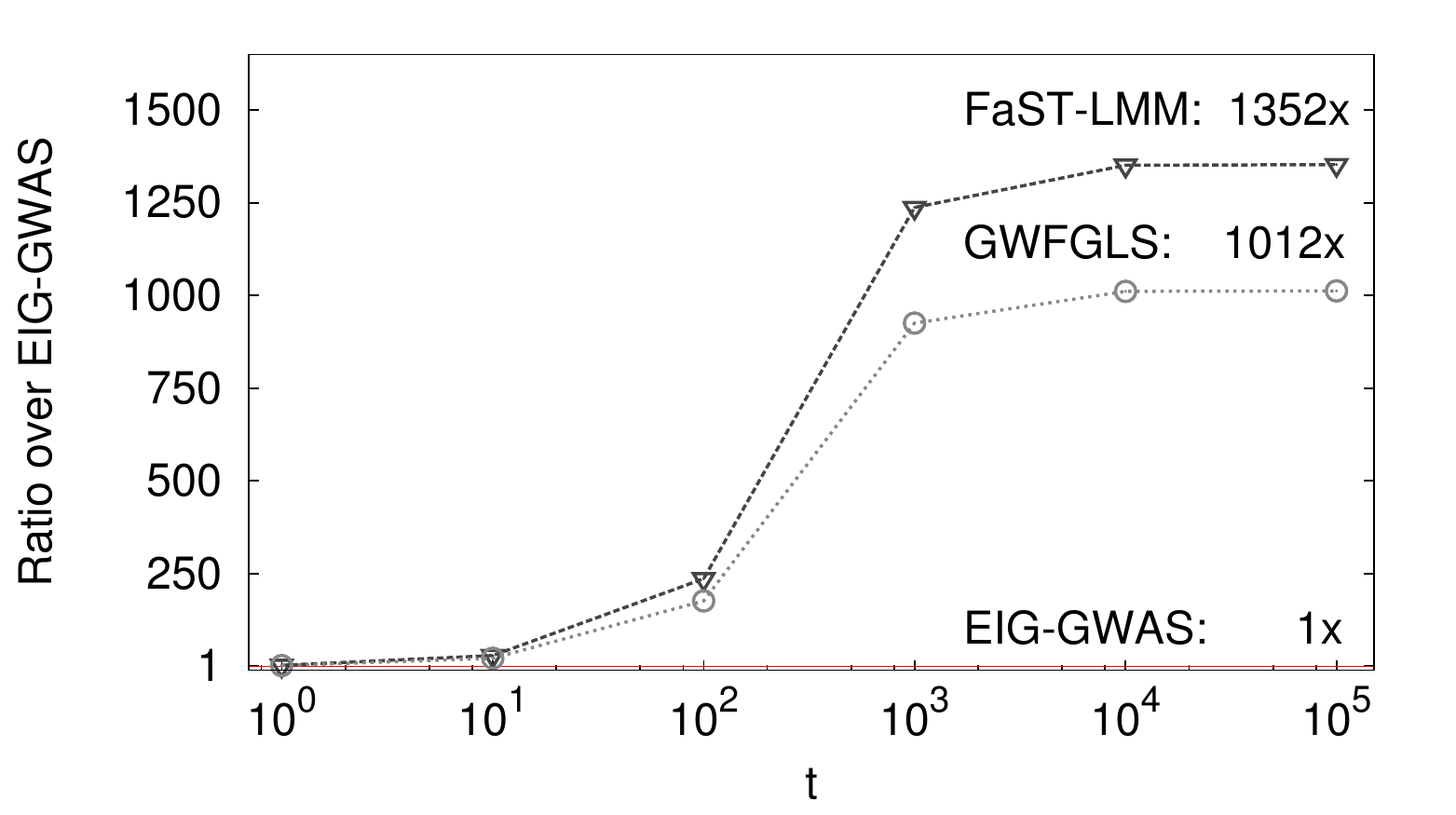}
\caption{
Solution of GWAS, Eq.~\eqref{eq:probDef}:
Execution time of {\sc fast-lmm} and {\sc gwfgls} over that of 
the compiler-generated {\sc eig-gwas}.
($n = 1000$, $p=4$, and $m = 10^6$; 40 cores.)}
\label{fig:gwas}
\end{minipage}
\hfill
\begin{minipage}[t]{.45\textwidth}
\centering
\includegraphics[scale=0.40]{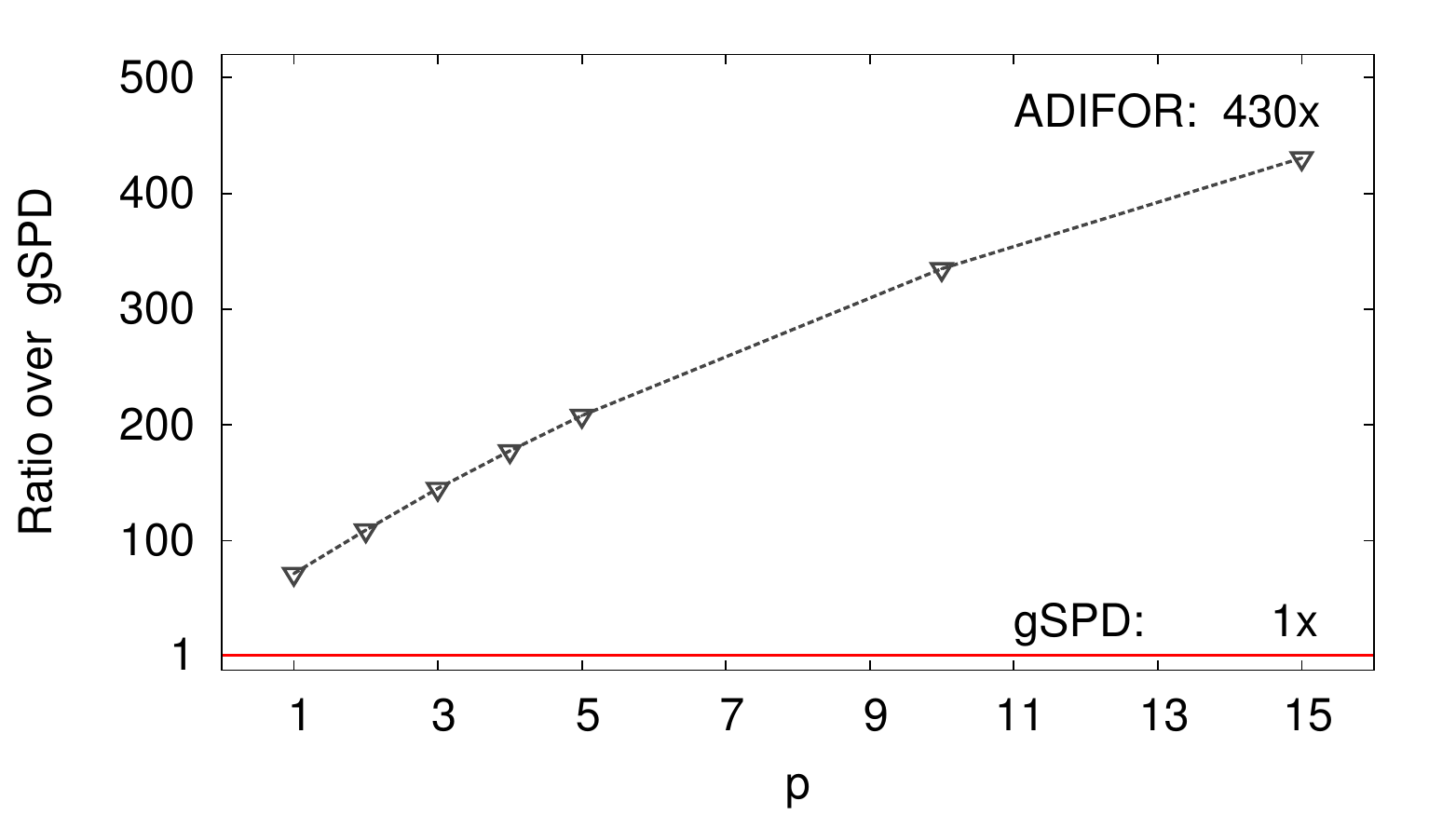}
\caption{Sensitivity analysis, Eq.~\eqref{eq:sens}:
  Execution time for the routine generated by ADIFOR over that of   
  the compiler-generated gSPD. 
  ($n = 1{,}000$; 1 core.)}
\label{fig:sensitivities}
\end{minipage}
\end{figure}

We first study the computation of the two-dimensional sequence of problems
arising as part of GWAS (Eq.~\eqref{eq:probDef}). We compare the performance of 
{\sc eig-gwas} (Algorithm~\ref{alg:alg-eigen}, tailored for sequences) 
with the two state-of-the-art libraries, 
{\sc gwfgls}, from the GenABEL project~\cite{genabel},
and {\sc fast-lmm}~\cite{Lippert2011}.
Figure~\ref{fig:gwas} displays the ratio for the execution time 
of these libraries over that of {\sc eig-gwas},
for increasing values of $t$. 
The results are impressive: {\sc eig-gwas} attains 1000-fold speedups.

Next, we present performance results for Eq.~\eqref{eq:sens}.
We compare the timings for gSPD (Algorithm~\ref{alg:many-sensitivities})
and the equivalent routine generated by the popular tool for sensitivity 
analysis---based on the automatic differentiation approach---ADIFOR~\cite{Bischof1992AGD,Bischof1996AAD}.
Figure~\ref{fig:sensitivities}, 
shows results for increasing $p$,
the length of the sequence; 
gSPD achieves speedups larger than 400.

Such remarkable results are justified by two factors,
the optimizations leading to an effective mapping onto building blocks, 
and the reduction of the computational cost due to the reuse of 
intermediate results across the sequences.

\section{Conclusions} \label{sec:conclusions}

We presented the design of a domain-specific compiler for linear algebra operations.
The compiler takes as input a matrix equation together with problem-specific knowledge,
and automatically generates a family of application-tailored algorithms.
The process centers around the decomposition of the equation 
into a series of calls to kernels provided by libraries such as BLAS and LAPACK. 
The decomposition is not unique, and even for simple equations many alternative
algorithms could be generated. In order to limit the output to competitive 
solutions, the compiler incorporates heuristics that aim at 
replicating---and at the same time extending---the
thought-process of a human expert. In this respect, we provided evidence 
that the compiler produces algorithms that match or even outperform those 
created by humans.

First, we detailed the application of the heuristics used by the compiler. 
The main idea behind them is to favor low-complexity algorithms. 
The benefits are two-fold: 
On the one hand, the heuristics allow the pruning of the search space;
on the other hand, they enable the generation of efficient algorithms tailored to the application.

Then, we uncovered the modules that constitute the compiler's engine.
One of the modules takes care of interfacing with the kernels provided
by numerical libraries. 
Two more modules are responsible for the management of knowledge, both
static and dynamic. As static knowledge, the compiler incorporates the
definition of operands and operators, together with their properties
and other algebraic identities.
Knowledge is also acquired dynamically via an inference engine capable
of deducing properties of matrix expressions.
These properties are central to the tuning of the algorithms: 
The more are available,
the better the tailoring.

We allow the input equation to be part of a sequence of related problems; 
in this case, the compiler makes use
of input information on the specific sequence to perform a data dependency analysis,
identify redundant computations, and reuse intermediate results.
We applied the compiler to equations arising as part of sensitivity and genome studies;
the produced algorithms exhibit, respectively, 100- and 1000-fold speedups.

\paragraph{\bf Future Work}  

There are many directions in which to extend and improve the compiler.
One of our most immediate objectives is the
support for implicit equations and higher-dimensional objects; these steps
require more advanced heuristics and optimizations.  On a different
front, in order to relieve the user from a tedious and error-prone
process, we plan to include the generation of C and Fortran code.

A challenging and critical component is the automatic selection of the
best algorithm. The mere operation count is not a reliable metric, so
we aim at incorporating advanced techniques for performance
prediction. A promising direction relies on a sample-based approach: 
The idea is to create performance models not for the competing algorithms, 
but only for those routines that are used as building blocks. 
By combining the models, it is then possible to make  
predictions and to rank the algorithms.~\cite{Peise2012:50}

\section{Acknowledgements} \label{sec:ack}
\sloppypar
The authors gratefully acknowledge the support received from the
Deutsche Forschungsgemeinschaft (German Research Association) through
grant GSC 111.

\bibliographystyle{splncs}
\bibliography{bibliography}

\begin{thebibliography}{10}

\bibitem{R}
{R Development Core Team}:
\newblock R: A Language and Environment for Statistical Computing.
\newblock R Foundation for Statistical Computing, Vienna, Austria. (2011)
  {ISBN} 3-900051-07-0.

\bibitem{10.1371/journal.pgen.1001256-short}
Lauc, G.,  et~al.:
\newblock {G}enomics {M}eets {G}lycomics--{T}he {F}irst {GWAS} {S}tudy of
  {H}uman {N}-{G}lycome {I}dentifies {HNF1}$\alpha$ as a {M}aster {R}egulator
  of {P}lasma {P}rotein {F}ucosylation.
\newblock PLoS Genetics \textbf{6}(12) (12 2010)  e1001256

\bibitem{Levy2009-short}
Levy, D.,  et~al.:
\newblock Genome-wide association study of blood pressure and hypertension.
\newblock Nature Genetics \textbf{41}(6) (Jun 2009)  677--687

\bibitem{Speliotes2010-short}
Speliotes, E.K.,  et~al.:
\newblock Association analyses of 249,796 individuals reveal 18 new loci
  associated with body mass index.
\newblock Nature Genetics \textbf{42}(11) (Nov 2010)  937--948

\bibitem{blas3}
Dongarra, J., Croz, J.D., Hammarling, S., Duff, I.S.:
\newblock A set of level 3 basic linear algebra subprograms.
\newblock ACM Trans. Math. Softw. \textbf{16}(1) (1990)  1--17

\bibitem{laug}
Anderson, E., Bai, Z., Bischof, C., Blackford, S., Demmel, J., Dongarra, J.,
  Du~Croz, J., Greenbaum, A., Hammarling, S., McKenney, A., Sorensen, D.:
\newblock {LAPACK} Users' Guide. Third edn.
\newblock Society for Industrial and Applied Mathematics, Philadelphia, PA
  (1999)

\bibitem{atlas-sc98}
Whaley, R.C., Dongarra, J.:
\newblock Automatically tuned linear algebra software.
\newblock In: SuperComputing 1998: High Performance Networking and Computing.
  (1998)

\bibitem{Bientinesi:2005:SDD}
Bientinesi, P., Gunnels, J.A., Myers, M.E., Quintana-Ort\'{i}, E.S., {v}an~de
  Geijn, R.A.:
\newblock The science of deriving dense linear algebra algorithms.
\newblock {ACM} Transactions on Mathematical Software \textbf{31}(1) (March
  2005)  1--26

\bibitem{Vuduc:2004uz}
Vuduc, R.W.:
\newblock Automatic performance tuning of sparse matrix kernels.
\newblock PhD thesis, University of California, Berkeley, CA, USA (January
  2004)

\bibitem{Baumgartner05synthesisof}
Baumgartner, G., Auer, A., Bernholdt, D.E., Bibireata, A., Choppella, V.,
  Cociorva, D., Gao, X., Harrison, R.J., Hirata, S., Krishnamoorthy, S.,
  Krishnan, S., chung Lam, C., Lu, Q., Nooijen, M., Pitzer, R.M., Ramanujam,
  J., Sadayappan, P., Sibiryakov, A., Bernholdt, D.E., Bibireata, A., Cociorva,
  D., Gao, X., Krishnamoorthy, S., Krishnan, S.:
\newblock Synthesis of high-performance parallel programs for a class of ab
  initio quantum chemistry models.
\newblock In: Proceedings of the IEEE. (2005)  2005

\bibitem{FFTW05}
Frigo, M., Johnson, S.G.:
\newblock The design and implementation of {FFTW3}.
\newblock Proceedings of the IEEE \textbf{93}(2) (2005)  216--231 Special issue
  on ``Program Generation, Optimization, and Platform Adaptation''.

\bibitem{Pueschel:05}
P{\"u}schel, M., Moura, J.M.F., Johnson, J., Padua, D., Veloso, M., Singer, B.,
  Xiong, J., Franchetti, F., Gacic, A., Voronenko, Y., Chen, K., Johnson, R.W.,
  Rizzolo, N.:
\newblock {SPIRAL}: Code generation for {DSP} transforms.
\newblock Proceedings of the IEEE, special issue on ``Program Generation,
  Optimization, and Adaptation'' \textbf{93}(2) (2005)  232-- 275

\bibitem{Logg:2010:DAF:1731022.1731030}
Logg, A., Wells, G.N.:
\newblock Dolfin: Automated finite element computing.
\newblock ACM Trans. Math. Softw. \textbf{37} (April 2010)  20:1--20:28

\bibitem{DxTer}
Marker, B., Poulson, J., Batory, D., van~de Geijn, R.:
\newblock Designing linear algebra algorithms by transformation: Mechanizing
  the expert developer.
\newblock In: Proceedings of VECPAR 2012. (2012) Accepted for publication.

\bibitem{Aho:86}
Aho, A.V., Sethi, R., Ullman, J.D.:
\newblock Compilers principles, techniques, and tools.
\newblock Addison-Wesley, Reading, MA (1986)

\bibitem{RECSY1}
Jonsson, I., {K\aa gstr\"om}, B.:
\newblock Recursive blocked algorithms for solving triangular systems---part i:
  one-sided and coupled sylvester-type matrix equations.
\newblock ACM Transactions on Mathematical Software \textbf{28}(4) (2002)
  392--415

\bibitem{RECSY2}
Jonsson, I., {K\aa gstr\"om}, B.:
\newblock Recursive blocked algorithms for solving triangular systems---part
  ii: Two-sided and generalized sylvester and lyapunov matrix equations.
\newblock ACM Transactions on Mathematical Software \textbf{28}(4) (2002)
  416--435

\bibitem{2002:USB:567806.567807}
Blackford, S., Demmel, J., Dongarra, J., Duff, I., Hammarling, S., Henry, G.,
  Heroux, M., Kaufman, L., Lumsdaine, A., Petitet, A., Pozo, R., Remington, K.,
  R., W.C.:
\newblock An updated set of basic linear algebra subprograms ({BLAS}).
\newblock ACM Trans. Math. Softw. \textbf{28}(2) (June 2002)  135--151

\bibitem{0021725}
Griewank, A., Walther, A.:
\newblock Evaluating derivatives - {P}rinciples and techniques of algorithmic
  differentiation (2. ed.).
\newblock SIAM (2008)

\bibitem{genabel}
Aulchenko, Y.S., Ripke, S., Isaacs, A., van Duijn, C.M.:
\newblock Genabel: an {R} library for genome-wide association analysis.
\newblock Bioinformatics \textbf{23}(10) (May 2007)  1294--6

\bibitem{Lippert2011}
Lippert, C., Listgarten, J., Liu, Y., Kadie, C.M., Davidson, R.I., Heckerman,
  D.:
\newblock Fast linear mixed models for genome-wide association studies.
\newblock Nat Meth \textbf{8}(10) (Oct 2011)  833--835

\bibitem{Bischof1992AGD}
Bischof, C.H., Carle, A., Corliss, G.F., Griewank, A., Hovland, P.D.:
\newblock {ADIFOR}: {G}enerating derivative codes from {F}ortran programs.
\newblock Scientific Programming \textbf{1}(1) (1992)  11--29

\bibitem{Bischof1996AAD}
Bischof, C.H., Carle, A., Khademi, P., Mauer, A.:
\newblock {ADIFOR} 2.0: Automatic differentiation of {F}ortran 77 programs.
\newblock IEEE Computational Science \& Engineering \textbf{3}(3) (1996)
  18--32

\bibitem{Peise2012:50}
Peise, E., Bientinesi, P.:
\newblock Performance modeling for dense linear algebra.
\newblock (November 2012) Accepted for publication in the proceedings of the
  3rd International Workshop on Performance Modeling, Benchmarking and
  Simulation of High Performance Computer Systems (PMBS12).

\end{thebibliography}

\end{document}